\documentclass[reprint,amsmath,amssymb]{revtex4-2}

\usepackage{tikz}
\usepackage{pgfplots}
\usepackage{circuitikz}
\usepackage[normalem]{ulem}
\usepackage{physics}

\usepackage{hyperref}
\usepackage{graphicx,amsmath,amsfonts}% Include figure files
\usepackage{amssymb}
\usepackage{dcolumn}% Align table columns on decimal point
\usepackage{mathrsfs}
\usepackage{xcolor}
\usepackage{bm}% bold math
\usepackage{enumerate}
\usepackage{pgfplots}
\usepackage{tikz}
\usetikzlibrary{decorations.markings}
\usetikzlibrary{plotmarks}
\usepackage[export]{adjustbox}

\usepackage{braket}
\usepackage{graphicx,amsmath,amsfonts}% Include figure files
\usepackage{dcolumn}% Align table columns on decimal point
\usepackage{mathrsfs}
\usepackage{xcolor}
\usepackage{calligra}
\usepackage{bm}% bold math
\usepackage{enumerate}
\usepackage{pgfplots}
\usepackage{tikz}
\usetikzlibrary{decorations.markings}
\usetikzlibrary{plotmarks}
\usepackage[export]{adjustbox}
\usepackage{stackengine}

\newcommand{\rb}{\mathbf{r}}

\makeatletter
\newsavebox{\@brx}
\newcommand{\llangle}[1][]{\savebox{\@brx}{\(\m@th{#1\langle}\)}%
  \mathopen{\copy\@brx\kern-0.5\wd\@brx\usebox{\@brx}}}
\newcommand{\rrangle}[1][]{\savebox{\@brx}{\(\m@th{#1\rangle}\)}%
  \mathclose{\copy\@brx\kern-0.5\wd\@brx\usebox{\@brx}}}

\colorlet{LightGray}{gray!10}
\makeatother

\definecolor{matlab1}{HTML}{0072BD}
\definecolor{matlab2}{HTML}{D95319}
\definecolor{matlab3}{rgb}{0.4940,0.1840,0.5560}

\newlength{\mywidth}
\newlength{\myheight}
\newlength{\mydelta}
\begin{document}
%\linenumbers % activate line numbering

\title{Spectral densities of a dispersive dielectric sphere \\ in the modified Langevin noise formalism}

%\title{Spectral densities of a dispersive electromagnetic environment out of equilibrium 
%in the modified Langevin noise formalism
%}

\author{Giovanni Miano, Loris Maria Cangemi, and Carlo Forestiere}
\email[]{carlo.forestiere@unina.it}
\affiliation{Department of Electrical Engineering and Information Technology, Universit\`{a} degli Studi di Napoli Federico II, via Claudio 21,  Napoli, 80125, Italy}

\begin{abstract}
This paper deals with the spectral densities of a dispersive dielectric object in the framework of macroscopic quantum electrodynamics based on the modified Langevin noise formalism. In this formalism, the electromagnetic field in the presence of a dielectric object has two contributions, one taking into account the polarization current fluctuations of the object and the other taking into account the vacuum field fluctuations scattered by the object. The combined effect of these fields on the dynamics of a quantum emitter is described via two independent continuous bosonic reservoirs, a medium-assisted reservoir and a scattering-assisted reservoir, each characterized by its own spectral density and initial quantum state. For initial thermal states of the two reservoirs at different temperatures, the standard approach based on the knowledge of the dyadic Green function of the dielectric object at the quantum emitter position cannot be employed.  We map the two reservoirs to a single equivalent reservoir with a temperature-dependent effective spectral density and initially in its vacuum state, focusing on the case of a homogeneous dielectric sphere. We derive analytical expressions for the medium-assisted, scattering-assisted, and effective spectral densities in this setting. We then study the dynamics of the quantum emitter for initial thermal states of the two reservoirs, adopting a non-perturbative approach.
\end{abstract}

\maketitle

\section{Introduction}
Harnessing the interaction between electromagnetic fields and matter is of growing importance across diverse areas of physics, from condensed matter physics to quantum optics and information. In particular, light–matter interaction in the strong-coupling regime enables the precise manipulation of the physical properties of hybrid systems and the preparation of nonclassical states of light (e.g. \cite{frisk_kockum_ultrastrong_2019,qin_quantum_2024,gonzalez_tudela_lightmatter_2024}).  Engineered coupling with electromagnetic resonators in cavity quantum electrodynamics platforms can profoundly alter the radiative properties of quantum emitters (e.g. \cite{chikkaraddy_single_molecule_2016,gonzalez_tudela_lightmatter_2024}), can regulate photochemical reaction rates (e.g. \cite{garcia-vidal_manipulating_2021,hsu_chemistry_2025}), and can control collective phenomena in quantum materials (e.g. \cite{schlawin_cavity_2022,jarc_cavity-mediated_2023}). The electromagnetic environments in these systems are open, dispersive, and absorbing, characteristics that pose challenges to field quantization. Macroscopic quantum electrodynamics offers a powerful phenomenological framework for quantizing the electromagnetic field in such complex systems, enabling a consistent description of light–matter interaction.

For spatially unbounded dispersive dielectrics, which fill the entire space, quantization can be carried out using the \textit{Langevin noise} formalism, where the electromagnetic field is generated by
the noise polarization currents of the dielectric according to the fluctuation-dissipation theorem (e.g. \cite{gruner_green-function_1996,dung_three-dimensional_1998,scheel_macroscopic_2008,westerberg_perturbative_2023}). For finite-size dielectrics, there is an additional contribution arising from the vacuum electromagnetic field fluctuations scattered by the dielectric, which cannot be disregarded (e.g. \cite{stefano_mode_2001,drezet_quantizing_2017,dorier_canonical_2019,dorier_critical_2020,forestiere_operative_2022,forestiere_integral_2023,na_numerical_2023,semin_canonical_2024}). The \textit{modified Langevin noise} formalism adds the scattering-assisted electromagnetic field to the medium-assisted electromagnetic field of the original Langevin noise model (e.g. \cite{stefano_mode_2001,drezet_quantizing_2017,na_numerical_2023}): it includes both the electromagnetic field generated by the medium fluctuations and the vacuum fluctuations on equal footing. The modified Langevin noise formalism has recently been justified starting from a phenomenological model and using the Heisenberg picture \cite{ciattoni_quantum_2024}. 
Consequently, in the framework of macroscopic quantum electrodynamics, the presence of dispersive dielectric objects of finite size involves the interaction of quantum emitters with two distinct and independent bosonic reservoirs, the medium-assisted reservoir and the scattering-assisted reservoir that can initially be in different quantum states \cite{miano_quantum_2025}.

The interaction of a quantum emitter with a bosonic reservoir initially in its vacuum quantum state depends only on the spectral density of the reservoir $\mathcal{J}(\omega)$, which encodes the density of states of the bosonic harmonic oscillators and the coupling between the emitter and the oscillators. In the regime of weak coupling, the spontaneous emission rate at the
emitter frequency $\omega_a$ is given by $2\pi \mathcal{J}(\omega_a)$ (e.g. \cite{h_p_breuer_and_f_petruccione_theory_2002,novotny_principles_2006,bassano_vacchini_open_2024}).
Recently, it has been shown that the modified Langevin noise formalism retrieves the expression\,$\mathcal{J}(\omega)=\frac{\omega^2}{\pi\hbar\varepsilon_0 c^2} \mathbf{p}\cdot \operatorname{Im}\left[\mathcal{G}_\omega\left(\mathbf{r}_a, \mathbf{r}_a\right)\right]\cdot\mathbf{p}$
for the spectral density of a dielectric object when both the medium-assisted and scattering-assisted bosonic reservoirs are initially in their vacuum quantum states \cite{na_numerical_2023,miano_quantum_2025},
 where $\mathcal{G}_\omega(\rb,\rb')$ is the dyadic Green function of the object, $\rb_a$ is the position vector of the emitter, and $\mathbf{p}$ is its transition dipole moment.

In Ref. \cite{miano_quantum_2025} we investigated the interaction of a quantum emitter with a dispersive dielectric object when the medium- and scattering-assisted reservoirs are initially in thermal states at the same temperature, focusing on a one-dimensional dielectric slab with the initial temperature of both reservoirs equal to zero. In this paper, we extend the treatment of Ref. \cite{miano_quantum_2025}, analyzing the interaction of a quantum emitter with a dispersive dielectric object when the medium-assisted reservoir and the scattering-assisted reservoir are initially in thermal quantum states at different temperatures, focusing on the case of a homogeneous dielectric sphere.    The scenario of two reservoirs initially in thermal quantum states with different temperatures may describe different experimental settings: for instance, cavities driven by thermal photon baths of interest in polaritonic chemistry (e.g., \cite{pannir-sivajothi_blackbody_2025,fassioli_controlling_2025}), and cryogenically cooled cavities (Fabry-Pérot,
nanoplasmonic structures, photonic crystal waveguides,
etc.) in the presence of thermal photons (e.g., \cite{jarc_cavity-mediated_2023,chiriaco_thermal_2024,bacciconi_dissipation_2025}).

 The quantum emitter interacts with two distinct reservoirs. Each reservoir is characterized by a proper spectral density and a proper initial quantum state. The medium-assisted spectral density $\mathcal{J}^{(M)}(\omega)$ arises exclusively from the medium-assisted electromagnetic field, while the scattering-assisted spectral density $\mathcal{J}^{(S)}(\omega)$ originates solely from the scattering-assisted electromagnetic field. The reduced dynamic of the quantum emitter depends on the individual spectral densities of the two reservoirs and the individual initial quantum states. Both spectral densities are connected to the dyadic Green function of the dielectric object and satisfy the fundamental relation $\mathcal{J}^{(M)}(\omega)+\mathcal{J}^{(S)}(\omega)=\frac{\omega^2}{\pi\hbar\varepsilon_0 c^2} \mathbf{p}\cdot \operatorname{Im}\left[\mathcal{G}_\omega\left(\mathbf{r}_a, \mathbf{r}_a\right)\right]\cdot\mathbf{p}$ \cite{ciattoni_quantum_2024}, \cite{miano_quantum_2025}.
Simply knowing the dyadic Green function of the dielectric object  at the position of the quantum emitter is insufficient to determine the individual spectral densities  and, therefore, the dynamics of the emitter. 

 The reduced dynamic of the quantum emitter can be evaluated by simulating its interaction with the two baths at different temperatures, or alternatively by introducing an equivalent single reservoir. Specifically, we show that the action of the two baths can be equivalently described by a zero-temperature bosonic reservoir \cite{tamascelli_efficient_2019}, with an effective spectral density given by the sum of the temperature-dependent medium-assisted and scattering-assisted spectral densities.
Then, exploiting this equivalence, we study the time evolution of a two-level quantum emitter coupled to a homogeneous dispersive dielectric sphere using a non-perturbative approach. We  analytically evaluate the medium-assisted and the scattering-assisted spectral densities of the dielectric sphere.

The paper is organized as follows. In Section II, we model the coupling of a quantum emitter with a dispersive dielectric object within the framework of the modified Langevin noise formalism. In Section III, we introduce the medium-assisted, scattering-assisted, and effective spectral densities of the dielectric object. In Section IV, we first discuss the behavior of these spectral densities for a homogeneous dielectric sphere; then, we analyze the evolution of a two-level quantum emitter for different scenarios of the initial temperatures of the medium-assisted and scattering-assisted baths, obtained using the matrix product state approach. In Section V, we give a summary and our conclusions. 

\section{Electromagnetic environment}

A quantum emitter interacts with a finite-size dielectric object in an unbounded space. We assume that the dielectric is linear, isotropic, and nondispersive in space; the dipole moment operator of the quantum emitter is given by $\hat{{\mu}}\mathbf{u}$ where $\hat{{\mu}}$ is the transition dipole moment operator of the emitter and $\mathbf{u}$ is a unit vector that indicates the direction of the dipole moment. We denote by $\rb_a$ the position vector of the emitter, by $V$ the region occupied by the dielectric object, and by $\varepsilon_\omega(\mathbf{r})$ its relative permittivity at frequency $\omega$.
The electromagnetic environment that surrounds the quantum emitter consists of both the polarization density field of the dielectric object and the electromagnetic field. 

The Hamiltonian of the composite system is
\begin{equation}
\hat{H} = \hat{H}_{A} + \hat{H}_{E} + \hat{H}_{I},
\end{equation}
where $\hat{H}_{A}$ is the bare emitter Hamiltonian, $\hat{H}_{E}$ is the bare Hamiltonian of the electromagnetic environment, and $\hat{H}_{I}$ is the interaction Hamiltonian. In the multipolar coupling scheme and within the dipole approximation $\hat{H}_{I}$ is given by
\begin{equation}
\label{eq:H_int}
\hat{H}_{I}=-\hat{\mu}\hat{V},
\end{equation}
where
\begin{equation}
\hat{V} = \mathbf{u} \cdot \left[\int_0^\infty d\omega \hat{\mathbf{E}}_\omega(\rb_a) + h.c.\right],
\end{equation}
and $\hat{\mathbf{E}}_\omega(\rb)$ is the monochromatic component of the electric field operator.

\subsection{Modified Langevin noise formalism}

In the modified Langevin noise formalism \cite{ciattoni_quantum_2024}, the electric field operator $\hat{\mathbf{E}}_\omega(\rb)$ has two contributions:
\begin{equation}
\label{eq:Elect}
    \hat{\mathbf{E}}_\omega(\rb) = \hat{\mathbf{E}}^{(M)}_{\omega} (\rb)+ \hat{\mathbf{E}}^{(S)}_{\omega} (\rb).
\end{equation}
The medium-assisted field $\hat{\mathbf{E}}^{(M)}_{\omega}$ is generated by the noise polarization currents of the dielectric object, while the scattering-assisted term $\hat{\mathbf{E}}^{(S)}_{\omega}$ takes into account the vacuum fluctuations scattered by the object.

The monochromatic component of the medium-assisted field $\hat{\mathbf{E}}_\omega^{(M)}$ is given by
\begin{equation}
\label{eq:Emed}
\hat{\mathbf{E}}_\omega^{(M)}(\mathbf{r})= \int_V d^3 \mathbf{r}^{\prime} \, \mathcal{G}_{e}\left(\mathbf{r}, \mathbf{r}^{\prime}; \omega\right) \cdot \hat{\mathbf{f}}_{\omega}\left(\mathbf{r}^{\prime}\right),
\end{equation}
where
\begin{equation}
\mathcal{G}_{e}\left(\mathbf{r}, \mathbf{r}^{\prime}; \omega\right)=i \frac{\omega^2}{c^2} \sqrt{\frac{\hbar}{\pi \varepsilon_0} \operatorname{Im}\left[\varepsilon_\omega\left(\mathbf{r}^{\prime}\right)\right]} \, \mathcal{G}_\omega\left(\mathbf{r}, \mathbf{r}^{\prime}\right),
\label{eq:GreenEq}
\end{equation}
$\mathcal{G}_\omega\left(\mathbf{r}, \mathbf{r}^{\prime}\right)$ is the dyadic Green function in the presence of the polarizable object at frequency $\omega$, $\hat{\mathbf{f}}_{\omega}\left(\mathbf{r}\right)$ is the monochromatic component of the bosonic field operator describing the polarization field at frequency $\omega$, whose support is the region $V$; $\varepsilon_0$ and $c$ are, respectively, the vacuum permittivity and the  speed of light in vacuum. 

The scattering-assisted field is expressed in terms of the scattering modes 
\begin{equation}
\mathbf{E}_{\omega \mathbf{n} \nu}(\mathbf{r}) =\sqrt{\frac{\hbar \omega^3}{16 \pi^3 \varepsilon_0 c^3}} \mathbf{F}_{\omega \mathbf{n} \nu}(\mathbf{r}),
\end{equation}
where $\mathbf{F}_{\omega \mathbf{n} \nu}(\mathbf{r})$ is the solution of the equation 
\begin{equation}
\left[\nabla\times \nabla\times-k_\omega^2 \varepsilon_\omega(\rb)\right] \mathbf{F}_{\omega \mathbf{n} \nu}=0,
\end{equation}
with the boundary condition
\begin{equation}
\mathbf{F}_{\omega \mathbf{n} \nu}(\mathbf{r}) \underset{r \rightarrow \infty}{\approx} e^{ik_{\omega}\rb\cdot \mathbf{n}}\mathbf{e}_{\mathbf{n} \nu};
\end{equation}
$k_\omega=\omega/c$, $\mathbf{n}$ is the unit vector along the wave vector $\mathbf{k} = k_{\omega}\mathbf{n}$ and $\mathbf{e}_{\mathbf{n}1}$, $\mathbf{e}_{\mathbf{n}2}$ are two mutually orthogonal polarization unit vectors that are orthogonal to $\mathbf{n}$.
The monochromatic component of the scattering-assisted field $\hat{\mathbf{E}}_\omega^{(S)}$ is given by
\begin{equation}
\label{eq:Escat}
\hat{\mathbf{E}}^{(S)}_\omega(\mathbf{r})=
\oint d o_{\mathbf{n}} \sum_\nu \mathbf{E}_{\omega \mathbf{n} \nu}(\mathbf{r}) \hat{g}_{\omega \mathbf{n} \nu},
\end{equation}
where $o_{\mathbf{n}}=(\theta_{\mathbf{n}},\phi_{\mathbf{n}})$ are the polar angles of the unit vector ${\mathbf{n}}$, $do_{\mathbf{n}}=\sin\theta_{\mathbf{n}} d\theta_{\mathbf{n}} d\phi_{\mathbf{n}}$ is the differential of the solid angle and the integration is performed over the whole solid angle with $\theta \in [0, \pi]$ and $\phi \in [0, 2 \pi]$. The operator $\hat{g}_{\omega \mathbf{n} \nu}$ is the monochromatic component of the bosonic operator at frequency $\omega$ that takes into account the vacuum fluctuations scattered by the dielectric object.

The bare Hamiltonian of the electromagnetic environment is given by
\begin{equation}\label{eq:HEtot}
\hat{H}_{E}= \hat{H}^{(M)}_E+ \hat{H}^{(S)}_E,
\end{equation}
where
\begin{equation}\label{eq:HEmedium}
\hat{H}^{(M)}_E=\int_0^{\infty} d \omega \hbar \omega \int_V d^3 \mathbf{r} \,\hat{\mathbf{f}}_{\omega}^{\dagger}(\rb) \cdot \hat{\mathbf{f}}_{\omega}(\rb),
\end{equation}
\begin{equation}\label{eq:HEscatt}
\hat{H}^{(S)}_E= \int_0^{\infty} d \omega \hbar \omega \oint d o_{\mathbf{n}} \sum_\nu \hat{g}_{\omega \mathbf{n} \nu}^{\dagger} \hat{g}_{\omega \mathbf{n} \nu}
\end{equation}
are, respectively, the contributions of the  medium-assisted and scattering-assisted fields. The interaction Hamiltonian is given by
\begin{equation}\label{eq:HItot}
\hat{H}_{I}=-\hat{\mu}\left[\hat{V}^{(M)} + \hat{V}^{(S)}\right],
\end{equation}
where $\hat{V}^{(M)}= \mathbf{u} \cdot \left[\int_0^\infty d\omega \hat{\mathbf{E}}_\omega^{(M)}(\rb_a) + h.c.\right]$ and $\hat{V}^{(S)}= \mathbf{u} \cdot \left[\int_0^\infty d\omega \hat{\mathbf{E}}_\omega^{(S)}(\rb_a) + h.c.\right]$ are, respectively, the contribution of the medium- and scattering-assisted fields.

The operators $\hat{\mathbf{f}}_{\omega}^{\dagger}$ and $\hat{\mathbf{f}}_{\omega}$ are the creation and annihilation operators of the medium-assisted excitations; the operators $\hat{g}_{\omega \mathbf{n} \nu}^{\dagger}$ and $\hat{g}_{\omega \mathbf{n} \nu}$ are the creation and annihilation operators of the scattering-assisted excitations. Any possible commutation relation between them vanishes, except the fundamental ones
\begin{equation}
{\left[\hat{\mathbf{f}}_{\omega}(\mathbf{r}), \hat{\mathbf{f}}_{\omega^{\prime}}^{\dagger}\left(\mathbf{r}^{\prime}\right)\right]} =\delta\left(\omega-\omega^{\prime}\right) \delta\left(\mathbf{r}-\mathbf{r}^{\prime}\right) \mathcal{I},
\end{equation}
\begin{equation}
{\left[\hat{g}_{\omega \mathbf{n} \nu}, \hat{g}_{\omega^{\prime} \mathbf{n}^{\prime} \nu^{\prime}}^{\dagger}\right] } =\delta\left(\omega-\omega^{\prime}\right) \delta\left(o_{\mathbf{n}}-o_{\mathbf{n}^{\prime}}\right) \delta_{\nu \nu^{\prime}},
\end{equation}
where $\mathcal{I}$ is the identity dyad and $\delta\left(o_{\mathbf{n}}-o_{\mathbf{n}^{\prime}}\right)=\delta\left(\theta_{\mathbf{n}}-\theta_{\mathbf{n}}^{\prime}\right) \delta\left(\varphi_{\mathbf{n}}-\varphi_{\mathbf{n}}^{\prime}\right) / \sin \theta_{\mathbf{n}}$. These commutation relations,  together with expressions \eqref{eq:Emed} and \eqref{eq:Escat}, guarantee the canonical commutation relations for the electromagnetic field and the matter field \cite{ciattoni_quantum_2024}. 

The fundamental integral identity \cite{ciattoni_quantum_2024}
\begin{equation}
\label{eq:sum}
\mathcal{M}_\omega(\rb,\rb')+ \mathcal{S}_\omega(\rb,\rb')=\frac{\hbar \omega^2}{\pi \varepsilon_0 c^2} \operatorname{Im}\left[\mathcal{G}_\omega\left(\mathbf{r}, \mathbf{r}^{\prime}\right)\right]
\end{equation}
holds, where 
\begin{equation}
    \mathcal{M}_\omega(\rb,\rb')=\int_V d^3 \mathbf{r''} \, \mathcal{G}_{e}(\mathbf{r}, \mathbf{r''; \omega}) \cdot \mathcal{G}_{e}^{* T}\left(\mathbf{r}^{\prime}, \mathbf{r''};\omega\right),
\end{equation}
and
\begin{equation}
    \mathcal{S}_\omega(\rb,\rb')=\oint d o_{\mathbf{n}} \sum_\nu \mathbf{E}_{\omega \mathbf{n} \nu}(\mathbf{r}) \mathbf{E}_{\omega \mathbf{n} \nu}^*\left(\mathbf{r}^{\prime}\right).
\end{equation}

The number of degrees of freedom of the electromagnetic environment can be reduced by applying the emitter-centered mode approach (e.g. \cite{buhmann_casimir-polder_2008}, \cite{feist_macroscopic_2021}) as in \cite{miano_quantum_2025}. For the two contributions to the bare Hamiltonian of the electromagnetic environment, we obtain (Appendix A). 
\begin{equation}
\hat{H}_{E}^{(M)}= \int_0^\infty d\omega \hbar \omega \hat{A}_\omega ^\dagger \hat{A}_\omega,
\end{equation}
\begin{equation}
\hat{H}_{E}^{(S)}= \int_0^\infty d\omega \hbar \omega \hat{B}_\omega ^\dagger \hat{B}_\omega,
\end{equation}
where $\{\hat{A}_\omega\}$ and $\{\hat{B}_\omega\}$ are two continuous sets of independent  bosonic annihilation operators relevant to the bright modes of the medium-assisted and scattering-assisted electromagnetic fields, i.e., the modes that effectively couple to the emitter. For the two contributions to the interaction operator $\hat{V}$, we obtain
\begin{equation}
\hat{V}^{(M)}=\left[ \int_0^\infty d\omega  g_{M}(\omega)\hat{A}_\omega +h.c.\right],
\end{equation}
\begin{equation}
\hat{V}^{(S)}=\left[ \int_0^\infty d\omega g_{S}(\omega)\hat{B}_\omega +h.c.\right],
\end{equation}
where
\begin{equation}
g_{M}(\omega)=\sqrt{ \int_V{d^3 \mathbf{r} \, \mathbf{u}\cdot [\mathcal{G}_{e}(\rb_a,\rb;\omega}) \cdot \mathcal{G}_{e}^{* T}(\rb_a,\rb;\omega) ] \cdot \mathbf{u}},
\label{eq:gM}
\end{equation}
\begin{equation}
g_{S}(\omega)=\sqrt{ \oint do_{\mathbf{n}} \, \mathbf{u} \cdot [\sum_\nu \mathbf{E}^*_{\omega \mathbf{n} \nu}(\mathbf{r}_a) \mathbf{E}_{\omega \mathbf{n} \nu}(\mathbf{r}_a)] \cdot \mathbf{u}}.
\label{eq:gS}
\end{equation}
In summary, the interaction of a quantum emitter with a dispersive dielectric object can be described by representing the electromagnetic environment through two independent continuous bosonic reservoirs, the medium-assisted bosonic reservoir and the scattering-assisted bosonic reservoir. The function $g_M(\omega)$ characterizes the interaction of the quantum emitter with the medium-assisted reservoir, and the function $g_S(\omega)$ characterizes the interaction with the scattering-assisted reservoir.

\section{Spectral densities}

The  dynamics of the emitter coupled to the two bosonic reservoirs can be solved numerically with a range of computational approaches \cite{schollwock_density-matrix_2011,medina_few-mode_2021}. However, this can be computationally intensive. In order to reduce the computational cost, we introduce an equivalent model based on a single surrogate reservoir in the same spirit as \cite{tamascelli_nonperturbative_2018,tamascelli_efficient_2019}.

\subsection{Correlation functions}

For initial product states of the entire system, $\hat{\rho}(0)=\hat{\rho}_{A0} \otimes \hat{\rho}^{(M)}_{E0}\otimes \hat{\rho}^{(S)}_{E0}$, where $\hat{\rho}_{A0}$, $\hat{\rho}^{(M)}_{E0}$, and $\hat{\rho}^{(S)}_{E0}$ are the initial density operators of the emitter, the medium-assisted bosonic reservoir, and the scattering-assisted bosonic reservoir, respectively, and for initial thermal states of the reservoirs, the reduced dynamic of the quantum emitter ${ \hat{\rho}}_A(t)=\text{Tr}_E[{{ \hat{\rho}}(t)}]$ depends only on its bare Hamiltonian $\hat{H}_A$, initial density operator $\hat{\rho}_{A0}$ and the correlation function of the electromagnetic environment $C(t)$ (e.g. \cite{h_p_breuer_and_f_petruccione_theory_2002}, \cite{bassano_vacchini_open_2024}) given by 
\begin{equation}
C(t) = \text{Tr}_E \left[\hat{U}^\dagger_E(t)\hat{V} \hat{U}_E(t) \hat{V}\hat{\rho}_{E0} \right],
\end{equation}
where $\hat{\rho}_{E0}=\hat{\rho}^{(M)}_{E0}\otimes\hat{\rho}^{(S)}_{E0}$ and 
\begin{equation}
\hat{U}_E(t)=\exp(-i\hat{H}_{E}t/\hbar)
\end{equation}
is the evolution operator of the electromagnetic environment in free evolution. This fundamental property allows us to design a single surrogate bosonic reservoir to compute the time evolution of the reduced density operator of the emitter. We denote by $T_M$ the initial temperature of the medium-assisted bosonic reservoir and by $T_S$ the initial temperature of the scattering-assisted bosonic reservoir. We obtain 
\begin{equation}
\label{eq:corr}
C(t) = C^{(M)}(t) + C^{(S)}(t),
\end{equation}
where 
\begin{equation}
\label{eq:corrM}
C^{(M)}(t)= \int_0^{\infty} \mathrm{d} \omega \, {\mathrm{g}}_{M}^2(\omega) \Theta(\omega t; \beta_{M} \hbar\omega),
\end{equation}
\begin{equation}
\label{eq:corrS}
C^{(S)}(t)= \int_0^{\infty} \mathrm{d} \omega \,{\mathrm{g}}_{S}^2(\omega) \Theta(\omega t; \beta_{S} \hbar\omega),
\end{equation}
with
\begin{equation}
\label{eq:tempfact}
\Theta(\omega t; \beta_{\alpha} \hbar\omega)=\operatorname{coth}\left(\frac{\beta_\alpha \hbar\omega}{2}\right) \cos (\omega \mathrm{t})-{i} \sin (\omega \mathrm{t}),
\end{equation}
$\beta_\alpha=1/(k_B T_\alpha)$ and $\alpha=M, S$.
 $C^{(M)}(t)$ is the contribution to the electromagnetic environment correlation function due to the medium-assisted bosonic bath at temperature $T_M$ and $C^{(S)}(t)$ is the contribution due to the scattering-assisted bosonic bath at temperature $T_S$. In summary, polarization fluctuations and vacuum fluctuations contribute in a different manner to the dynamics of the quantum emitter, because functions $g_M(\omega)$ and $g_S(\omega)$ are different and the temperatures of the two reservoirs can be different.
 
 \subsection{Medium- and scattering- assisted spectral densities}
 The temperatures of the two reservoirs and the functions $g_M^2(\omega)$ and $g_S^2(\omega)$ characterize the correlation functions $C^{(M)}(t)$ and $C^{(S)}(t)$. It is convenient to introduce the \textit{medium-assisted spectral density} $\mathcal{J}^{(M)}(\omega)=\left(\frac{\mu}{\hbar}\right)^2g_M^2 (\omega)$ and the \textit{scattering-assisted spectral density} $\mathcal{J}^{(S)}(\omega)=\left(\frac{\mu}{\hbar}\right)^2g_S^2 (\omega)$, both defined for $0\leq\omega<\infty$, where $\mu$ is the transition dipole moment of the quantum emitter. They are not independent; in fact, we have as a consequence of \eqref{eq:sum}
\begin{equation}
\label{eq:gequiv}
\mathcal{J}^{(M)}(\omega)    +    \mathcal{J}^{(S)}(\omega) =\frac{1}{2\pi}\Gamma(\omega;\rb_a),
\end{equation}
where
\begin{equation}
\label{eq:gequiv1}
\Gamma(\omega;\rb_a) = \frac{2\omega^2}{\hbar\varepsilon_0 c^2} \mathbf{p}\cdot \operatorname{Im}\left[\mathcal{G}_\omega\left(\mathbf{r}_a, \mathbf{r}_a\right)\right]\cdot\mathbf{p} 
\end{equation}
and $\mathbf{p}=\mu \mathbf{u}$.
This relation has a very important physical meaning according to the Poynting theorem. Consider the classical electromagnetic field generated by an electric dipole  oscillating at frequency $\omega$, with dipole moment $ \bf{p}$ and position vector ${\bf{r}}_a$ in the presence of the dielectric body.  The quantity $\frac{\pi}{2}\hbar \omega \mathcal{J}^{(M)}(\omega)$ coincides with the average electromagnetic power absorbed by the dielectric body, while the quantity $\frac{\pi}{2}\hbar \omega \mathcal{J}^{(S)}(\omega)$ coincides with the average electromagnetic power radiated towards infinity. Their sum $(\frac{1}{4}\hbar \omega)\Gamma(\omega;\rb_a)$ is equal to the average electromagnetic power emitted by the oscillating electric dipole according to the Poynting vector. 

Since the relative permittivity $\varepsilon_\omega$ tends to $1$ for $\omega \rightarrow \infty$, the medium-assisted spectral density tends to $0$ for $\omega \rightarrow \infty$, while the scattering-assisted spectral density tends to that of vacuum, which diverges as $\omega^3$. As a consequence, the integral \eqref{eq:corrS} diverges. This divergence is due to the dipole approximation in the expression of the interaction Hamiltonian \eqref{eq:H_int}, which is only valid in the long-wavelength limit. This inconsistency can be resolved by introducing a cutoff frequency in the integral \eqref{eq:corrS} to eliminate the interaction of the emitter with the high-frequency modes or by removing the dipole approximation. In this paper, we introduce a cutoff frequency.

The quantity $\Gamma(\omega;\rb_a)$ is the spontaneous emission rate of the quantum emitter in the  weak coupling limit when both baths are initially in the vacuum state, according to Fermi's golden rule. In fact, when the temperature of both baths is equal to zero, the time evolution of the reduced density operator of the quantum emitter can be described by introducing an equivalent bosonic reservoir initially in the vacuum quantum state and with effective spectral density $\mathcal{J}(\omega)=\mathcal{J}^{(M)}(\omega)+\mathcal{J}^{(S)}(\omega)=\Gamma(\omega;\rb_a)/2\pi$. In the literature based on the Langevin noise formalism, a similar relation is widely used; however, the conditions under which it remains valid are not always clearly stated. When the temperatures of the two baths are not equal to zero, this result  is no longer valid, as already pointed out in \cite{miano_quantum_2025}.

\subsection{Effective spectral density}

To extend this equivalence criterion to baths  at different temperatures, we proceed as in \cite{tamascelli_efficient_2019}. We extend the integrals in \eqref{eq:corrM} and in \eqref{eq:corrS} to the whole real frequency axis and introduce the extended spectral densities $\mathcal{J}^{(M)}_{ext}(\omega)$ and $\mathcal{J}^{(S)}_{ext}(\omega)$ with support  over the whole real frequency axis, which read
\begin{equation}
\label{eq:spectr}
  \mathcal{J}^{(\alpha)}_{ext}(\omega)=
  \text{sign}(\omega)\mathcal{J}^{(\alpha)}(|\omega|),  
\end{equation}
with $\alpha=M, S$. It is immediate that $C^{(\alpha)} (t)$ with $\alpha=M, S$ can be rewritten as
\begin{equation}
\label{eq:corr2}
C^{(\alpha)}(t)= \left(\frac{\hbar}{\mu}\right)^2 \int_{
-\infty}^{+\infty} \mathrm{d} \omega \mathcal{J}^{(\alpha)}_{td}(\omega)e^{-i\omega t},
\end{equation}
where we have introduced the temperature-dependent spectral density
\begin{equation}
\mathcal{J}^{(\alpha)}_{td}(\omega)= \frac{\mathcal{J}^{(\alpha)}_{ext} (\omega)}{2}\left[1+\coth\left(\frac{\beta_{\alpha}\hbar\omega}{2}\right)\right].
\end{equation}
Therefore, the correlation function $C(t)$ of the electromagnetic environment can be expressed as
\begin{equation}
C(t)= \left(\frac{\hbar}{\mu}\right)^2\int_{
-\infty}^{+\infty} \mathrm{d} \omega \mathcal{J}_{eff}(\omega) e^{-i\omega t},
\end{equation}
where
\begin{equation}
\label{eq:equiv}
    \mathcal{J}_{eff}(\omega)= \mathcal{J}^{(M)}_{td}(\omega)+\mathcal{J}_{td}^{(S)}(\omega).
\end{equation}
The function $\mathcal{J}^{(M)}_{td}$ is the temperature-dependent spectral density of the medium-assisted bath and $\mathcal{J}^{(S)}_{td}$ is the temperature-dependent spectral density of the scattering-assisted bath, which take into account the effects of thermal excitations. The effective spectral density $\mathcal{J}_{eff}(\omega)$ of the electromagnetic environment is non-negative for any frequency value and depends on the temperatures of the two baths. 

In conclusion, when the initial temperatures of the two baths are different from zero, the dynamics of the reduced density operator of the quantum emitter can also be described by a single surrogate bosonic reservoir initially in the vacuum state, with positive and negative frequencies, characterized by the effective spectral density $\mathcal{J}_{eff}(\omega)$. The bare Hamiltonian of the surrogate bosonic reservoir is given by
\begin{equation}\label{eq:EqHam}
\hat{H}_E^{(eff)}=\int_{-\infty}^{+\infty} d\omega \, \hbar\omega \, \hat{a}_\omega^\dagger \hat{a}_\omega,
\end{equation}
where $\hat{a}_\omega^\dagger$ and $\hat{a}_\omega$ are creation and annihilation operators, and the interaction Hamiltonian is given by
\begin{equation}\label{eq:intHam}
\hat{H}_I^{(eff)}=-\hat{\mu} \left(\frac{\hbar}{\mu}\right) \int_{-\infty}^{+\infty} d\omega \sqrt{\mathcal{J}_{eff}(\omega)}(\hat{a}_\omega^\dagger+\hat{a}_\omega).
\end{equation}
When the temperatures of the two baths are equal, $\beta_M=\beta_S=\beta_0$, combining expressions \eqref{eq:equiv}, \eqref{eq:gequiv} and \eqref{eq:spectr} we obtain
\begin{multline}
\label{eq:spectreq}
\mathcal{J}_{eff}(\omega)=\frac{1}{4\pi}\text{sign}(\omega)\left[1+\coth\left(\frac{\beta_{0}\hbar\omega}{2}\right)\right] \Gamma(\omega;\rb_a) \,.
\end{multline}
When the temperatures tend to zero, this relation reduces to $\mathcal{J}_{eff}(\omega)=\frac{1}{2\pi}\theta(\omega)\Gamma(\omega;\rb_a)$ where $\theta(\omega)$ is the Heaviside function.
 The single surrogate bosonic reservoir, by construction, provides the reduced dynamic of the quantum emitter, but it is not able to describe the dynamics of the electromagnetic environment composed of the medium-assisted and scattering-assisted reservoirs. Once the two-time correlation functions of the observables describing the quantum emitter have been evaluated, the expectation values of the observables describing the electromagnetic environment can be computed from the time evolution of the operators $\hat{A}_\omega$ and $\hat{B}_\omega$ in the Heisenberg picture, which can be  evaluated analytically.

\section{Homogeneous dielectric sphere}

The findings presented in the preceding two sections are valid regardless of the shape of the dielectric object. In this Section, we apply them to a quantum emitter interacting with a homogeneous dielectric sphere, with the emitter positioned outside the sphere. We denote by $a$ the radius of the sphere and by $d$ the distance of the quantum emitter from the center of the sphere. We consider both the Drude and Debye models to describe the dielectric permittivity of the sphere. In the Drude model, the relative dielectric permittivity is given by $\varepsilon_\omega=[1-\omega_p^2/(\omega^2+i\nu\omega)]$ where $\omega_p$ is the plasma frequency and $\nu$ is the relaxation frequency. For the Debye model, $\varepsilon_\omega$ is given by $\varepsilon_\omega=[1+\chi_0/(1-i\tau\omega)]$ where $\chi_0$ is the static susceptibility value and $\tau$ is the relaxation time.

We first analyze the behavior of the medium-assisted spectral density $\mathcal{J}^{(M)}(\omega)$ and the scattering-assisted spectral density $\mathcal{J}^{(S)}(\omega)$ and compare them with the vacuum spectral density $\mathcal{J}_{vac}(\omega)$. 
The analytical expressions of $\mathcal{J}^{(M)}(\omega)$ and $\mathcal{J}^{(S)}(\omega)$ are given in Appendix B. The vacuum spectral density is given by 
\begin{equation}
\nonumber
  \mathcal{J}_{vac}(\omega)=(k_{\omega}/k_{r})^3 \mathcal{J}_0,  
\end{equation}
 where 
 \begin{equation}
\nonumber
\mathcal{J}_0 = \frac{1}{6\pi^2} \frac{k_r}{\hbar\, \varepsilon_0} \left({\mu k_r}\right)^2,
 \end{equation}
 $k_r$ is a reference wavenumber given by
 \begin{equation}
 \nonumber
     \quad \; k_r = 
\begin{cases}
			k_p=\omega_p/c & \text{Drude,}\\
            k_c=\omega_c/c & \text{Debye,}
		 \end{cases}
 \end{equation}
and $\omega_c=c/(a\sqrt{\chi_0})$. By definition, the characteristic spectral density $\mathcal{J}_0$ is equal to the value of the vacuum spectral density at the reference wavenumber $k_r$, indicating the coupling strength between the quantum emitter and the vacuum field.

Then, we examine the behavior of the effective spectral density $\mathcal{J}_{eff}(\omega)$ for different initial thermal quantum states of the electromagnetic environment. Eventually, we investigate the time evolution of a two-level quantum emitter coupled to a Drude dielectric sphere using the matrix product state approach.

\subsection{Medium- and scattering-assisted spectral densities}

\begin{figure}
    \centering
    \includegraphics[width=\linewidth]{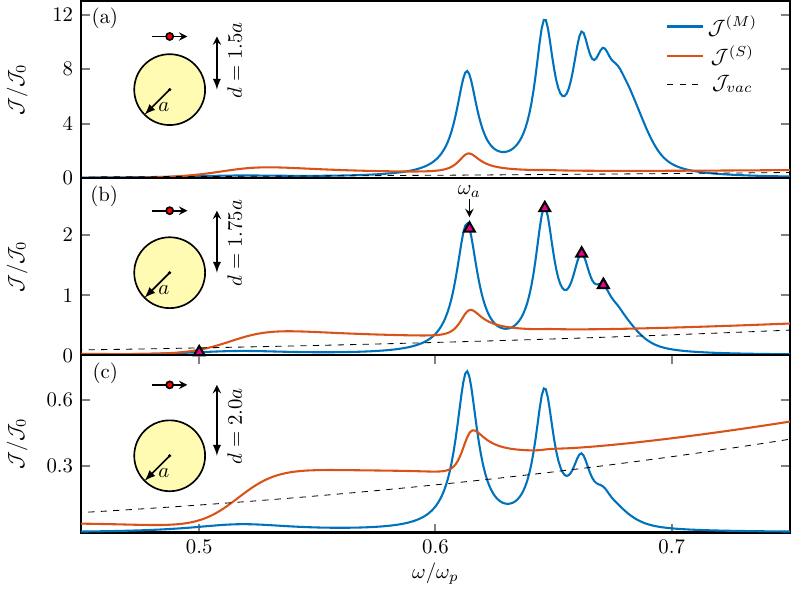}
    \caption{Drude sphere: normalized spectral densities $\mathcal{J}^{(S)}/\mathcal{J}_0$, $\mathcal{J}^{(M)}/\mathcal{J}_0$ and $\mathcal{J}_{vac}/\mathcal{J}_0=(\omega/\omega_p)^3$ versus $\omega/\omega_p$ with $k_pa=1$, $\nu/\omega_p = 0.01$ and (a) $d/a=1.5$, (b) $d/a=1.75$, (c) $d/a=2$, where $k_p=\omega_p/c$ and $\mathcal{J}_0 = \frac{1}{6 \pi^2} \frac{k_p}{\hbar\, \varepsilon_0 } \left({\mu k_p}\right)^2$. The triangular markers denote the resonance positions (as determined by Eq. \eqref{eq:PlasmonicRes} in Appendix \ref{sec:Resonances}), which, from left to right, correspond to the electric dipole, quadrupole, octupole, hexadecapole, and dotriacontapole \textit{plasmonic} modes.}
    \label{fig:DrudeJ}
\end{figure}

\begin{figure}
    \centering
    \includegraphics[width=\linewidth]{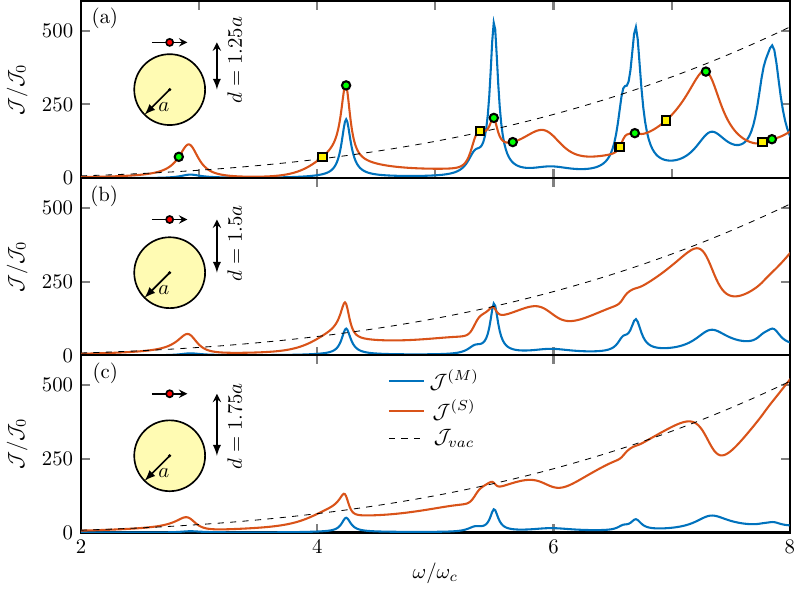}
    \caption{Debye sphere: normalized spectral densities $\mathcal{J}^{(S)}/\mathcal{J}_0$, $\mathcal{J}^{(M)}/\mathcal{J}_0$ and $\mathcal{J}_{vac}/\mathcal{J}_0=(\omega/\omega_c)^3$ versus $\omega/\omega_c$ with $\chi_0=15$, $\tau \omega_c=0.01 {/\sqrt{\chi_0}}$ and (a) $d/a=1.25$, (b) $d/a=1.50$, (c) $d/a=1.75$, where $\omega_c=c/(a\sqrt{\chi_0})$, $k_c=\omega_c/c$ and $\mathcal{J}_0 = \frac{1}{6 \pi^2} \frac{k_c}{\hbar\, \varepsilon_0 } \left({\mu k_c}\right)^2$. Green circle markers denote the resonance positions of \textit{dielectric} modes of magnetic type, given by \eqref{eq:DielResTE} in Appendix \ref{sec:Resonances}, which, from left to right correspond to the magnetic dipole $(n,\ell)=(1,1)$, m. quadrupole $(2,1)$, m. octupole $(3,1)$, second-order m. dipole $(1,2)$, m. hexadecapole $(4,1)$, second-order m. quadrupole $(2,2)$, m. dotriacontapole $(5,1)$. 
    Yellow square markers denote the resonance positions of \textit{dielectric} modes of electric type, given by \eqref{eq:DielResTM}) in Appendix \ref{sec:Resonances}, which, from left to right, correspond to the toroidal dipole $(n,\ell)=(1,1)$, t. quadrupole $(n,\ell)=(2,1)$, t. octupole $(n,\ell)=(3,1)$, second-order t. dipole $(n,\ell)=(1,2)$, t. hexadecapole $(n,\ell)=(4,1)$.}
    \label{fig:DebyeJ}
\end{figure}

We present the behavior of $\mathcal{J}^{(M)}(\omega)$ and $\mathcal{J}^{(S)}(\omega)$ when the orientation of the dipole moment is tangential to the sphere, as shown in Figures \ref{fig:DrudeJ} and \ref{fig:DebyeJ}. We find qualitatively similar results also for the radial orientation.

Figure \ref{fig:DrudeJ} shows $\mathcal{J}^{(M)}/\mathcal{J}_0$, $\mathcal{J}^{(S)}/\mathcal{J}_0$, and $\mathcal{J}_{vac}/\mathcal{J}_0=(\omega/\omega_p)^3$ versus $\omega/\omega_p$ for the Drude dielectric with $k_pa=1$, $\nu/\omega_p=0.01$ and (a) $d/a=1.50$, (b) $d/a=1.75$, and (c) $d/a=2.00$. These cases are relevant to the metallic spheres used in nanoplasmonics \cite{maier_plasmonics_2007}, whose electromagnetic response is dominated by plasmonic modes. Figure \ref{fig:DebyeJ} shows $\mathcal{J}^{(M)}/\mathcal{J}_0$, $\mathcal{J}^{(S)}/\mathcal{J}_0$, and $\mathcal{J}_{vac}/\mathcal{J}_0=(\omega/\omega_c)^3$ versus $\omega/\omega_c$ for the Debye dielectric with $\chi_0=15$ ($k_ca=0.25$), $\tau \omega_c=0.01 {/\sqrt{\chi_0}}$ and (a) $d/a=1.25$, (b) $d/a=1.50$, and (c) $d/a=1.75$. These cases are relevant to high permittivity spheres such as silicon used in nanophotonics \cite{krasnok_all-dielectric_2012}, whose electromagnetic response is dominated by dielectric modes \cite{forestiere_resonance_2020}. 
As expected, the curves of the medium-assisted and scattering-assisted spectral densities show peaks that are correlated with the resonance frequencies of the electromagnetic modes of the sphere, which have been evaluated using the analytical expressions given in Appendix \ref{sec:Resonances}.
The spectral density $\mathcal{J}^{(M)}$ tends linearly to zero as the frequency tends to zero, instead $\mathcal{J}^{(S)}$ tends to zero faster than the vacuum spectral density $\mathcal{J}_{vac}$. When the frequency tends to infinity, $\mathcal{J}^{(M)}$ tends rapidly to zero and $\mathcal{J}^{(S)}$ increases as the vacuum spectral density $\mathcal{J}_{vac}$.
The order of magnitude of $\mathcal{J}^{(M)}$ decreases rapidly as $d/a$ increases, while the order of magnitude of $\mathcal{J}^{(S)}$ does not change substantially.
The amplitudes of $\mathcal{J}^{(M)}$ and $\mathcal{J}^{(S)}$ can be very large compared to the amplitude of $\mathcal{J}_{vac}$ near the resonance peaks.

For the investigated Drude sphere, the amplitude of $\mathcal{J}^{(S)}$ exceeds that of $\mathcal{J}^{(M)}$ in the low-frequency range near the dipolar plasmonic resonance, as illustrated in Fig. \ref{fig:DrudeJ}. However, when the emitter is sufficiently close to the sphere and at frequencies located near higher-order plasmonic resonances, such as the electric quadrupole, octupole, hexadecapole, and dotriacontapole, the amplitude of $\mathcal{J}^{(M)}$ becomes significantly larger than that of $\mathcal{J}^{(S)}$.
The bandwidths of these plasmonic modes are limited by a combination of radiative and material losses, as detailed in Appendix \ref{sec:Resonances}. For the electric dipole mode of the studied sphere (with $k_pa = 1$), the bandwidth is predominantly governed by radiative losses and thus remains  approximately independent of the ratio $\nu/\omega_p$. Conversely, higher-order resonance peaks exhibit narrower bandwidths dominated by material dissipation, which increases with $\nu/\omega_p$. Furthermore, the peak amplitudes of the medium-assisted spectral density increase as either $d/a$ or $k_p a$ decrease.

In the case of the Debye sphere, when the quantum emitter is sufficiently near the sphere, there are frequency intervals in which the amplitude of $\mathcal{J}^{(M)}$ is higher than the amplitude of $\mathcal{J}^{(S)}$ and vice versa, as illustrated in Fig. \ref{fig:DebyeJ}.  Similar to the plasmonic case, the bandwidths of the dielectric modes are determined by both radiative and material losses, following the expression given in Appendix \ref{sec:Resonances}. For the magnetic dipole mode of the studied sphere (with $k_c a = 0.25$) both types of losses contribute significantly. In contrast, the resonance peaks of higher-order modes with $\ell = 1$ are narrower, primarily limited by  material losses, and their bandwidth increases with $\tau \omega_c $. For dielectric modes with $\ell = 2$, the radiative and material losses are generally comparable. The peak amplitudes of the medium-assisted spectral density grow with increasing $\tau\omega_c$.

\subsection{ Effective spectral density}\label{sec:extendspectral}

We now examine the behavior of the effective spectral density for different temperatures of the medium-assisted bath and the scattering-assisted bath in scenarios where the order of magnitude of $\mathcal{J}^{(M)}/\mathcal{J}_0$ is comparable to the order of magnitude of $\mathcal{J}^{(S)}/\mathcal{J}_0$. In particular, we consider as scenario I the case of Fig. \ref{fig:DrudeJ}(b), namely $k_pa=1$, $\nu/\omega_p = 0.01 $, and $d/a=1.75$. As scenario II we considered the case shown in Fig. \ref{fig:DebyeJ}(b), namely $\chi_0=15$, $\tau \omega_c=0.01 { /\sqrt{\chi_0}}$ and $d/a=1.50$.

Since the amplitude of the medium-assisted spectral density tends to zero linearly as the frequency tends to zero, the temperature-dependent spectral density $\mathcal{J}_{td}^{(M)}$ is continuous at $\omega=0$ and its value is  nonzero. The amplitude of $\mathcal{J}_{td}^{(S)}$ tends to zero faster than the vacuum spectral density as the frequency tends to zero, hence $\mathcal{J}_{td}^{(S)}(\omega=0)=0$.

\begin{figure}
    \centering
    \includegraphics[width=\linewidth]{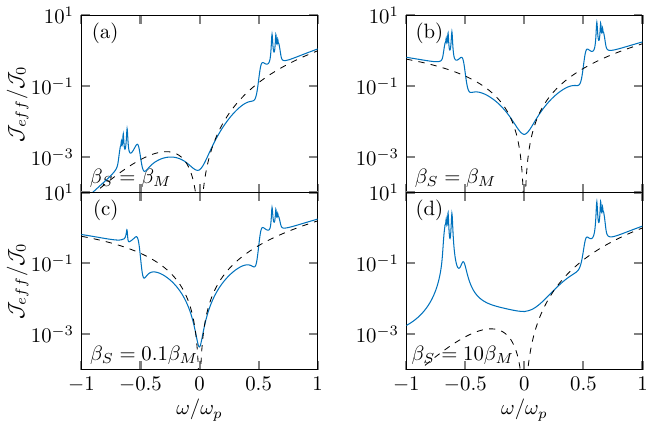}
    \caption{Drude sphere: effective spectral density $\mathcal{J}_{eff}$ normalized to $\mathcal{J}_0$ versus $\omega/\omega_p$ for $d/a=1.75$, $k_pa=1$, $\nu/\omega_p=0.01$, and different values of the inverse temperature parameters $\beta_M$ and $\beta_S$:
    (a) $\beta_M =\beta_S=10/(\hbar \omega_p)$; (b) $\beta_M =\beta_S=1/(\hbar \omega_p)$;
    (c) $\beta_M =10\beta_S=10/(\hbar \omega_p)$; 
    (d) $\beta_S=10\beta_M =10/(\hbar \omega_p)$. 
    The dashed lines represent the normalized temperature-dependent spectral density
    $\frac{\mathcal{J}_{vac}(\omega)}{2}[1+\coth\left(\frac{\beta_{S}\hbar\omega}{2}\right)]/\mathcal{J}_0$.}
    \label{fig:EffJDrude}
\end{figure}

\begin{figure}
    \centering
    \includegraphics[width=\linewidth]{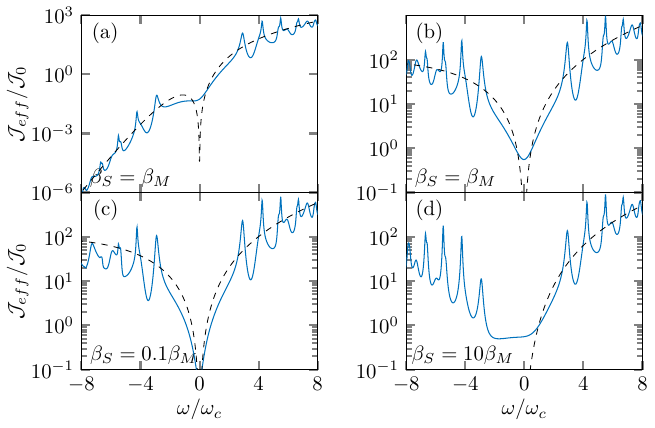}
    \caption{Debye sphere: effective spectral density $\mathcal{J}_{eff}$ normalized to $\mathcal{J}_0$ versus $\omega/\omega_c$ for $d/a=1.50$, $\chi_0=15$, $\tau \omega_c=0.01{ /\sqrt{\chi_0}}$, different values of the inverse temperature parameters $\beta_M$ and $\beta_S$: (a) $\beta_M =\beta_S =2.5/(\hbar \omega_c)$; (b) $\beta_M =\beta_S =0.25/(\hbar \omega_c)$; (c) $\beta_M =10\beta_S/(\hbar \omega_c) =2.5$; (d) $\beta_S=10\beta_M=2.5/(\hbar \omega_c)$. 
    The dashed lines represent  the normalized temperature-dependent spectral density
    $\frac{\mathcal{J}_{vac}(\omega)}{2}[1+\coth\left(\frac{\beta_{S} \hbar \omega}{2}\right)]/\mathcal{J}_0$.}
    \label{fig:EffJDebye}
\end{figure}

Figure \ref{fig:EffJDrude} shows the behavior of the effective spectral density $\mathcal{J}_{eff}$ normalized to $\mathcal{J}_0$ versus $\omega/\omega_p$ for a Drude sphere with $d/a=1.75$, $k_pa=1$, $\nu/\omega_p=0.01$, and different values of the inverse temperature parameters $\beta_M$ and $\beta_S$. Figure 4 shows the behavior of $\mathcal{J}_{eff}/\mathcal{J}_0$ versus $\omega/\omega_c$ for a Debye sphere with $d/a=1.50$, $\chi_0=15$ ($k_ca=0.25)$, $\tau \omega_c=0.01 /\sqrt{\chi_0}$ and different values of the temperature parameters $\beta_M$ and $\beta_S$. The dashed lines represent the normalized temperature-dependent vacuum spectral density $\frac{\mathcal{J}_{vac}(\omega)}{2}[1+\coth\left(\frac{\beta_{S} \hbar \omega}{2}\right)]/\mathcal{J}_0$.

We first comment on the cases in which the two baths
have equal temperatures. When the temperatures are close to zero, e.g., $\beta_M=\beta_S=10/(\hbar \omega_p)$, Fig. \ref{fig:EffJDrude}(a), and $\beta_M =\beta_S =2.5/(\hbar \omega_c)$, Fig. \ref{fig:EffJDebye}(a), the amplitudes of the effective spectral densities are  nearly  zero for negative frequencies, as expected. When the temperatures are high, for example, $\beta_M =\beta_S =1/(\hbar \omega_p)$ Fig. \ref{fig:EffJDrude}(b) and $\beta_M =\beta_S =0.25/(\hbar \omega_c)$, Fig. \ref{fig:EffJDebye}(b), the amplitudes of the effective spectral densities at negative frequencies are comparable  to  the amplitudes at positive frequencies, as expected. 

We now consider the cases in which the temperatures of the two baths are different. 
When the temperature of the scattering-assisted bath is ten times the temperature of the medium-assisted bath, e.g. $\beta_M =10\beta_S=10/(\hbar \omega_p)$, Fig. \ref{fig:EffJDrude}(c), and $\beta_M=10\beta_S=2.5/(\hbar \omega_c)$, Fig. \ref{fig:EffJDebye}(c), the amplitude of the effective spectral density is of the same order of magnitude of the amplitude of the vacuum spectral density at the same temperature, at both positive and negative frequencies. Conversely, when the temperature of the medium-assisted bath is ten times the temperature of the scattering-assisted bath, e.g. $\beta_S=10\beta_M =10/(\hbar \omega_p)$, Fig. \ref{fig:EffJDrude}(d), and $\beta_S=10\beta_M =2.5/(\hbar \omega_c)$, Fig. \ref{fig:EffJDebye}(d), the amplitude of the effective spectral densities deviates significantly from the amplitude of the vacuum spectral density for negative frequencies.

\subsection{Time evolution of a two-level quantum emitter}\label{subsec:emitter}

We now consider a two-level quantum emitter in the presence of a Drude dielectric sphere with $d/a=1.75$, $k_pa=1$ and $\nu/\omega_p=0.01$, and study its non-equilibrium dynamics. We investigate its dynamics using the equivalent model based on the single bosonic reservoir with Hamiltonian $\hat{H}_E^{(eff)}$ given by \eqref{eq:EqHam}, initially in the vacuum quantum state and with the temperature-dependent spectral density $\mathcal{J}_{eff}(\omega)$ given by \eqref{eq:equiv}.

We denote the Pauli matrices by $\hat{\sigma}_i$,
with $i=x, y, z$,
and the eigenstates of $\hat{\sigma}_z$ by ${\ket{\pm}}$, that is, $\hat{\sigma}_{z}\ket{\pm}=\pm\ket{\pm}$. The bare Hamiltonian of the two-level quantum emitter reads $\hat{H}_{A}={\hbar\omega_a}(\hat{\sigma}_{z}/2)$ where $\omega_a$ is the bare transition frequency. The electric dipole moment operator is given by $\hat{\mu} = \mu \hat{\sigma}_{x}$ where $\mu$ is the transition dipole moment, thus the interaction Hamiltonian is $
\hat{H}_I^{(eff)}=-\hbar\hat{\sigma}_{x} \int_{-\infty}^{+\infty} d\omega \sqrt{ \mathcal{J}_{eff}(\omega)}(\hat{a}_\omega^\dagger+\hat{a}_\omega)$. The resulting interaction Hamiltonian has the form of the quantum Rabi model in the presence of continuum reservoirs \cite{forn2017ultrastrong,frisk_kockum_ultrastrong_2019}. 

The emitter is initially prepared in the pure state $\hat{\rho}_{A}(0)=\ketbra{x-}$ where $\ket{x-}=(1/\sqrt{2}) (\ket{+}-\ket{-})$ is an eigenstate of $\hat{\sigma}_x$. We choose the bare transition frequency of the emitter equal to the resonance frequency of the electric quadrupole mode, $\omega_a/\omega_p=0.6144$. We found similar results for values of $\omega_a/\omega_p$ in the frequency range that  includes the resonance frequencies of the leading plasmonic modes.

We simulated the composite system with Hamiltonian $\hat{H}^{(eff)}=\hat{H}_A+\hat{H}_E^{(eff)}+\hat{H}_I^{(eff)}$ using the matrix product states (MPS) approach, with $N_{bm}^{+}=200$ discrete positive frequency bosonic modes and $N_{bm}^{-}=200$ discrete negative frequency bosonic modes, and a maximum local dimension of $N_{\text{ph}}=2$, Appendix \ref{sec:DetMPS}. This allows us to simulate non-perturbative effects \cite{peropadre2013nonequilibrium,forn2017ultrastrong,RyuMPS2023} in the dynamics of the whole system \cite{schollwock_density-matrix_2011,prior_efficient_2010,   weimer_simulation_2021}.
Although this approach may limit the spectral frequency range of the baths to be investigated, as well as the final time of the simulation, it provides a direct method to study the nonequilibrium dynamics of the emitter and to avoid the limitations imposed by pseudomode-based techniques \cite{lednev_lindblad_2024}.

Quite recently, the model in Equations \eqref{eq:EqHam} and \eqref{eq:intHam} has been used as a starting point for  the time-evolving density operator with orthogonal polynomials (TEDOPA) approach \cite{tamascelli_efficient_2019}. However, compared to TEDOPA, we do not perform the mapping to a chain Hamiltonian \cite{prior_efficient_2010,  tamascelli_efficient_2019}. Instead, we simulate the time evolution of the initial product state of the emitter and the equivalent bosonic reservoir and then compute the expectation values of the Pauli operators describing the emitter at different points in time. 

\begin{figure}
    \centering \includegraphics[width=\linewidth]{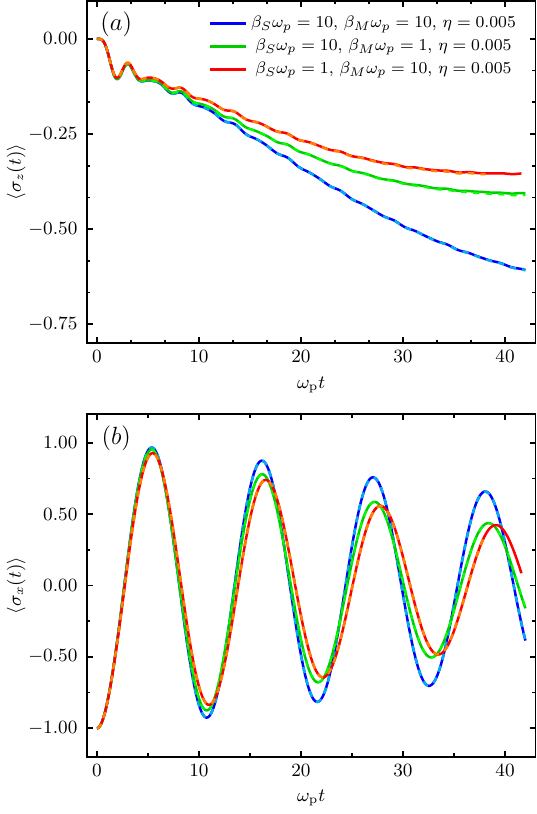}
    \caption{Expectation values of $\ev{\sigma_z(t)}$ (a) and  $\ev{\sigma_x(t)}$ (b) computed by means of the MPS approach for $\omega_{a}=\omega_2$, where $\omega_2$ is the resonance frequency of the quadrupole plasmonic mode of the Drude sphere, $\eta=\mathcal{J}_0/\omega_p= 5\cdot 10^{-3}$, $\omega_{\text{cut}}=3\omega_p$, and different values of inverse temperatures $\beta_{S}\omega_{\text{p}}=\beta_{M}\omega_{\text{p}}=10$, $\beta_{S}\omega_{\text{p}}=1,\beta_{M}\omega_{\text{p}}=10$, and $\beta_{S}\omega_{\text{p}}=10, \beta_{M}\omega_{\text{p}}=10$. Continuous (dashed) lines denote the choice of the local dimension of the oscillator Fock space $N_{\text{ph}}=1$ ($N_{\text{ph}}=2$).  }
    \label{fig:szMPSbetaS10betaS10}
\end{figure}

In Figure \ref{fig:szMPSbetaS10betaS10} we plot the time evolution of the expectation values $\ev{\sigma_{z}(t)}$ and $\ev{\sigma_{x}(t)}$ for the coupling strength $\eta=\mathcal{J}_0/\omega_p=5\cdot 10^{-3}$ and different values of the inverse temperatures of the baths. The reduced state dynamics undergoes a damped behavior in time, marked by intervals of non-monotonic behavior.
%Notice that the long time recurrences in the population dynamics are due to the nature of the interaction. Indeed, it has the form of the quantum Rabi model; 
Nonmonotonic behavior of the populations of the emitter state signals non-Markovian effects occurring at weak coupling strengths. 
At long times, the reduced state of the emitter is determined by the properties of both spectral densities and by their respective temperatures and cannot be predicted on the basis of the knowledge of the damping rates in the Born-Markov approximation $\gamma(\omega)$ (see Appendix D). This is because off-resonance processes, which lie beyond the validity of both the Markov and rotating wave approximations, can play a significant role. In particular, the scattering-assisted bath exerts a much stronger influence on the interaction dynamics compared to the  medium-assisted one, owing to the distinct shapes of the spectral functions $\mathcal{J}^{(S)}(\omega)$ and $\mathcal{J}^{(M)}(\omega)$. It follows that for $\beta_{S}< \beta_{M}$, the population of the reduced state of the emitter in the excited state is greater than the ground state one, even if the damping rates obey $\gamma_{M}(\omega_{a})>\gamma_{S}(\omega_{a})$. On the other hand, for the same value of the coupling strength $\eta$, the coherence of the emitter state tends to decay following typical dissipative behavior, where the decay times and frequency depend on the values of the reservoir temperatures. 
This feature is reminiscent of the physics of general spin-boson models \cite{weissbook,LeggettSBM87,HUR2008,Carrega2015heat,chen2025suppression}, where the reduced state of the emitter at long times depends both on the coupling strength $\eta$ and on the bath temperatures.   

\begin{figure}
    \centering %\includegraphics[width=\linewidth]{Figures/SzMPSweak.pdf}
    \includegraphics[width=\linewidth]{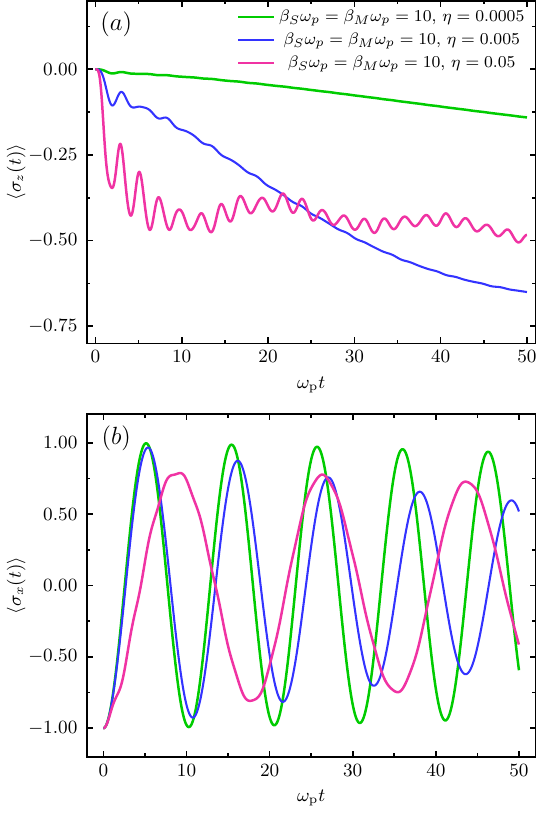}
   \caption{Expectation values of $\ev{\sigma_z(t)}$ (a) and $\ev{\sigma_x(t)}$  (b) for $\omega_{a}=\omega_2$, where $\omega_2$ is the resonance frequency of the quadrupole plasmonic mode of the Drude sphere, with $\omega_{\text{cut}}=3\omega_p$, computed by means of the MPS approach for three different values of the adimensional coupling strength $\eta=\mathcal{J}_0/\omega_p=5\cdot 10^{-4}, 5\cdot 10^{-3}, 5\cdot 10^{-1}$. The local dimension of the Fock space has been set to $N_{\text{ph}}=2$.}
\label{fig:szMPSbetaS1betaM11}
\end{figure}

The effects of correlations between the emitter and many bosonic modes of the two different baths inherently change the dynamics as the coupling strength $\eta$ increases. In the case of light-matter interactions, correlations lead to Rabi oscillations \cite{frisk_kockum_ultrastrong_2019}, that is, an amount of energy is exchanged back and forth between the baths and the emitter. This is confirmed by the time evolution of $\ev{\sigma_{z}(t)}$ and $\ev{\sigma_{x}(t)}$ shown in Fig. \ref{fig:szMPSbetaS1betaM11} obtained with three different values of coupling strength $\eta$, for fixed temperatures of the two reservoirs. Note that, as the coupling $\eta$ becomes stronger, one observes the simultaneous emergence of low-frequency oscillatory behavior and faster energy exchange with the reservoirs, which are not present in the previous regime. This feature is also evident from the coherence, which persists even at long times, oscillating with a frequency smaller than the bare emitter frequency, so that the emitter does not show the conventional dissipative behavior anymore. This also has several consequences on the energy fluxes between the two different baths and on the spectrum of the emitted radiation, which deserve further analysis.
  
%\begin{figure}
%    \centering \includegraphics[width=\linewidth]{Figures/SzMPSsca10m1.pdf}
%    \caption{}
%     \label{fig:szMPSbetaS10betaM1}
%\end{figure}

%\begin{figure}
%    \centering \includegraphics[width=\linewidth]{Figures/SzMPSsca1m10.pdf}
%    \caption{}
%\label{fig:szMPSbetaS1betaM10}
%\end{figure}

\section{Conclusions and outlook}

The spectral density encodes the relevant information that governs the interaction of a quantum emitter with a dispersive dielectric object, playing a central role in determining the dynamics of the emitter. Within the framework of macroscopic quantum electrodynamics based on the modified Langevin noise formalism, the emitter couples to two independent bosonic reservoirs: a medium-assisted reservoir describing the fluctuations of the polarization current density of the dielectric and a scattering-assisted reservoir describing the fluctuation of the vacuum field scattered by the object. 

This paper investigates the interaction of a quantum emitter with these two bosonic reservoirs, assuming an initial product state in which the reservoirs are prepared in  thermal state at different temperatures.
We show that each bosonic reservoir is characterized by its own spectral density. The medium-assisted spectral density is proportional to the average electromagnetic power absorbed by the dielectric object when the expectation value of the dipole moment of the quantum emitter oscillates at frequency $\omega$. The scattering-assisted spectral density is proportional to the average electromagnetic power radiated toward infinity. The reduced dynamic of the quantum emitter depends on the individual spectral densities of the two reservoirs and the individual initial quantum states.

We further introduced an \textit{effective} spectral density, defined as the sum of the temperature-dependent medium- and scattering-assisted spectral densities, allowing the dynamics of the emitter to be described through an equivalent single zero-temperature bosonic reservoir. This mapping facilitates a non-perturbative analysis of the dynamics of the emitter.

We apply this framework to the case of a homogeneous dielectric sphere, providing analytical expressions for the medium-assisted and scattering-assisted spectral densities. We numerically investigate them as a function of the physical parameters of the system using the Drude and Debye models for the permittivity. Our results reveal that the spectral densities encode the resonant electromagnetic response of the sphere, which strongly depends on geometric parameters and material dispersion. 
Then, using the effective spectral density, we investigated the non-Markovian time evolution of a two-level emitter coupled to a Drude sphere via the matrix product state approach.

Our simulations demonstrate that differences in the initial temperatures of the two reservoirs significantly influence the relaxation and decoherence dynamics of the emitter, highlighting the combined effects of scattering-assisted and medium-assisted fields. Furthermore, we observed that as the coupling strength increases, a coherent energy exchange between the emitter and the reservoirs emerges, accompanied by persistent coherence and low-frequency oscillations. These phenomena are characteristic of generalized spin-boson models and coherent Rabi dynamics. Our findings thus underline the inadequacy of simplified descriptions based solely on the values of the dyadic Green functions, damping rates derived within the Born–Markov and rotating wave approximations, emphasizing the necessity of fully characterizing the spectral structure of the electromagnetic environment.

The results presented in this paper can be extended to any magnetodielectric material that can be described by a linear constitutive relation. Future work will focus on extending this framework to more general dielectric geometries  using standard techniques from classical computational electromagnetics (e.g., finite element methods or boundary element methods), to multiple quantum emitters, and on cases where the bosonic reservoirs are not initially in thermal states.

\begin{acknowledgements}
This work was supported by the Italian Ministry of University and Research (MUR) through the PNRR Project PE0000023-NQSTI (C.F. and G.M.) and through the PNRR Project No. CN00000013-ICSC (L.M.C.)
\end{acknowledgements}

\appendix

\section{Reduced Hamiltonian}\label{sec:APPA}

For the problem at hand, the number of degrees of freedom of the electromagnetic environment can be reduced by applying the emitter-centered mode approach (e.g., \cite{buhmann_casimir-polder_2008}, \cite{feist_macroscopic_2021}) as in \cite{miano_quantum_2025}.

We start with the representation of the field operator $\hat{\mathbf{f}}_{\omega}(\rb)$. We consider the monochromatic operator $\hat{A}_{\omega}$ defined as
\begin{equation}
\hat{A}_{\omega}=\int_{V}d^3\mathbf{r}  \,  \boldsymbol{\alpha}_{\omega}(\rb)\cdot \hat{\mathbf{f}}_{\omega}(\mathbf{r}),
\end{equation}
where
\begin{equation}
\boldsymbol{\alpha}_{\omega}(\rb)=\frac{\mathbf{u} \cdot \mathcal{G}_{e}(\mathbf{r}_a, \mathbf{r}; \omega)}{g_M(\omega)}
\end{equation}
and $g_M(\omega)$ is an arbitrary normalization parameter. We choose $g_M(\omega)$ in such a way that the commutator between $\hat{A}_{\omega}$ and $\hat{A}_{\omega}^\dagger$ is
\begin{equation}
{\left[\hat{A}_{\omega}, \hat{A}_{\omega^{\prime}}^{\dagger}\right] } =\delta\left(\omega-\omega^{\prime}\right),
\end{equation}
and obtain
\begin{equation}
g_M(\omega)=\sqrt{ \int_V{d^3 \mathbf{r} \, \mathbf{u}\cdot [\mathcal{G}_{e}(\rb_a,\rb; \omega}) \cdot \mathcal{G}_{e}^{* T}(\rb_a,\rb; \omega) ] \cdot \mathbf{u}} \,.
\end{equation}
Then, the contribution of the medium-assisted field to the interaction operator $\hat{V}$ is given by
\begin{equation}
\hat{V}^{(M)}= \left[ \int_0^\infty d\omega g_M(\omega) \hat{A}_\omega +h.c. \right].
\end{equation}
On the other hand, we can always express the field operator $\hat{\mathbf{f}}_{\omega}(\mathbf{r})$ as
\begin{equation}
\hat{\mathbf{f}}_{\omega}(\mathbf{r})=\boldsymbol{\alpha}_{\omega}^*(\rb) \hat{A}_\omega + \sum_m [\boldsymbol{\alpha}_{\omega}^m(\rb)]^*\hat{A}_{\omega}^m,
\end{equation}
where the orthonormal set of vector fields $\{\boldsymbol{\alpha}_{\omega}^m(\rb)\}$ spans the subspace orthogonal to $\boldsymbol{\alpha}_{\omega}(\rb)$, that is, $ \int_Vd^3\rb [\boldsymbol{\alpha}_{\omega}^{m}(\rb)]^* \cdot \boldsymbol{\alpha}_{\omega}(\rb) = 0 $. Note that each $\boldsymbol{\alpha}_{\omega}^m(\rb)$ does not couple to the emitter; $\hat{A}_\omega$ is the emitter-centered bright mode of the medium-assisted field, while $\{\hat{A}_{\omega}^m\}$ are an infinite number of dark modes. Then, the contribution of the medium-assisted electromagnetic field to $\hat{H}_{E}$ is given by
\begin{equation}
\hat{H}_{E}^{(M)}= \int_0^\infty d\omega \hbar \omega \hat{A}_\omega ^\dagger \hat{A}_\omega +  \int_0^\infty d\omega \hbar \omega \sum_m (\hat{A}_{\omega}^{m})^\dagger \hat{A}_{\omega}^m.
\end{equation}

We now consider the representation of $\hat{g}_{\omega \mathbf{n} \nu}$. We introduce the monochromatic operator $\hat{B}_{\omega}$ defined by
\begin{equation}
\hat{B}_{\omega}=\oint d o_{\mathbf{n}} \sum _\nu \, {\beta}_{\omega \mathbf{n} \nu } \hat{g}_{\omega \mathbf{n} \nu },
\end{equation}
where
\begin{equation}
{\beta}_{\omega \mathbf{n} \nu } =\frac{\mathbf{u} \cdot \mathbf{E}_{\omega \mathbf{n} \nu}(\mathbf{r}_a)}{g_S(\omega)}.
\end{equation}
Here, $g_S(\omega)$ is an arbitrary real normalization  parameter chosen such that the commutator relation
\begin{equation}
{\left[\hat{B}_{\omega}, \hat{B}_{\omega^{\prime}}^{\dagger}\right] } =\delta\left(\omega-\omega^{\prime}\right)
\end{equation}
holds. Thus, we obtain for $g_S(\omega)$
\begin{equation}
g_S(\omega)=\sqrt{ \oint do_{\mathbf{n}} \, \mathbf{u} \cdot [\sum_\nu \mathbf{E}^*_{\omega \mathbf{n} \nu}(\mathbf{r}_a) \mathbf{E}_{\omega \mathbf{n} \nu}(\mathbf{r}_a)] \cdot \mathbf{u}} \,.
\end{equation}
Then, the contribution of the scattering-assisted field to $\hat{V}$ is given by
\begin{equation}
\hat{V}^{(S)}=\left[ \int_0^\infty d\omega g_S(\omega)\hat{B}_\omega +H.c.\right].
\end{equation}
On the other hand, the field operator $\hat{g}_{\omega \mathbf{n} \nu}$ can always be expressed as
\begin{equation}
\hat{g}_{\omega \mathbf{n} \nu}= {\beta}_{\omega \mathbf{n} \nu }^* \hat{B}_\omega + \sum_m [{\beta}_{\omega \mathbf{n} \nu}^m]^* \hat{B}_{\omega}^m,
\end{equation}
where  $\{{\beta}_{\omega \mathbf{n} \nu}^m\}$ is 
an orthonormal set of vector fields spanning the subspace orthogonal to ${\beta}_{\omega \mathbf{n} \nu }$, that is, $\int do_{\mathbf{n}} \sum_\nu [{\beta}_{\omega \mathbf{n} \nu}^m]^* {\beta}_{\omega \mathbf{n} \nu} = 0 $. Note that every ${\beta}_{\omega \mathbf{n} \nu}^m$ does not couple to the emitter; $\hat{B}_\omega$ is the emitter-centered bright mode of the scattering-assisted field, and $\{\hat{B}_{\omega}^m\}$ are an infinite number of dark modes. Consequently, the contribution of the scattering-assisted field to $\hat{H}_{E}$ is expressed as
\begin{equation}
\hat{H}_{E}^{(S)}= \int_0^\infty d\omega \hbar \omega \hat{B}_\omega ^\dagger \hat{B}_\omega +  \int_0^\infty d\omega \hbar \omega \sum_m (\hat{B}_{\omega}^m)^\dagger \hat{B}_{\omega}^m.
\end{equation}
Since dark modes are
decoupled from the rest of the system, they do not contribute to the reduced dynamics of the quantum emitter. As a consequence, we can set
\begin{equation}
\hat{H}_{E}^{(M)}= \int_0^\infty d\omega \hbar \omega \hat{A}_\omega ^\dagger \hat{A}_\omega,
\end{equation}
\begin{equation}
\hat{H}_{E}^{(S)}= \int_0^\infty d\omega \hbar \omega \hat{B}_\omega ^\dagger \hat{B}_\omega.
\end{equation}

\section{Analytical expressions for the spectral densities $\mathcal{J}_M(\omega)$ and $\mathcal{J}_S(\omega)$}

We have introduced the characteristic spectral density $\mathcal{J}_0 = \frac{1}{6 \pi^2} \frac{k_r}{\hbar\, \varepsilon_0} \left({\mu k_r}\right)^2$ where $k_r$ is a reference wavenumber that is equal to $k_p=\omega_p/c$ for the Drude sphere and to $k_c=\omega_c/a$ with $\omega_c=c/(a\sqrt{\chi_0})$ for the Debye sphere.

The scattering-assisted spectral density $\mathcal{J}^{(S)}(\omega)=\left(\frac{\mu}{\hbar}\right)^2 g_S^2(\omega)$, with $g_S$ defined in Eq. \eqref{eq:gS}, can be derived by evaluating the power radiated to infinity by an electric dipole located near a spherical surface. Kim, Leung, and George \cite{kim_classical_1988} analytically computed this power by determining the flux of the Poynting vector through a spherical surface at infinity. The normalized spectral densities for the two cases, when the dipole orientation is tangential or radial to the spherical surface and located at a distance $d$ from the sphere center, are given as follows:
 \begin{multline}
\frac{\mathcal{J}^{(S)}_\parallel \left( \omega, d \right)}{\mathcal{J}_0}=\frac{3}{4} \left( \frac{k_\omega}{k_r} \right)^3 \sum_{n=1}^{\infty}(2 n+1) \times 
\\ \left[\left|j_n(k_\omega d)+A_n h_n^{(1)}(k_\omega d)\right|^2+\left|\frac{\psi_n^{\prime}(k_\omega d)+B_n \zeta_n^{\prime}(k_\omega d)}{k_\omega d}\right|^2\right], 
\end{multline}

\begin{multline}
\frac{\mathcal{J}^{(S)}_\perp \left( \omega, d \right)}{\mathcal{J}_0} = \frac{3}{2}  \left( \frac{k_\omega}{k_r} \right)^3 \sum_{n=1}^{\infty} n(n+1)(2 n+1) \times \\ \left|\frac{j_n(k_\omega d)+B_n h_n^{(1)}(k_\omega d)}{k_\omega d}\right|^2,
\end{multline}
where $j_n$ are spherical Bessel functions, $h_n^{(1)}$ are spherical Hankel functions of the first kind, $\psi_n (x) = x j_n(x)$ and $\zeta_n(x) = x h_n^{(1)}(x) $ are the Riccati-Bessel functions \cite{abramowitz_handbook_1948,bohren_absorption_2008}.
The Mie coefficients $A_n$ and $B_n$ that appear above are explicitly defined by: 

\begin{equation}
\begin{aligned}
A_n & =\frac{j_n\left(k_\omega a\right) \psi_n^{\prime}\left(k_\omega \sqrt{\varepsilon_\omega} a\right)- j_n\left(k_\omega  \sqrt{\varepsilon_\omega} a\right) \psi_n^{\prime}\left(k_\omega a\right)}{j_n\left(k_\omega \sqrt{\varepsilon_\omega} a\right) \zeta_n^{\prime}\left(k_\omega a\right)-h_n^{(1)}\left(k_\omega a\right) \psi_n^{\prime}\left(k_\omega \sqrt{\varepsilon_\omega} a\right)}, \\
B_n & =\frac{ j_n\left(k_\omega a\right) \psi_n^{\prime}\left(k_\omega \sqrt{\varepsilon_\omega} a\right)-\varepsilon_\omega j_n\left(k_\omega \sqrt{\varepsilon_\omega} a\right) \psi_n^{\prime}\left(k_\omega a\right)}{\varepsilon_\omega j_n\left(k_\omega \sqrt{\varepsilon_\omega} a\right) \zeta_n^{\prime}\left(k_\omega a\right)- h_n^{(1)}\left(k_\omega a\right) \psi_n^{\prime}\left(k_\omega \sqrt{\varepsilon_\omega} a\right)}.
\end{aligned}
\end{equation}

Similarly, the medium-assisted spectral density $\mathcal{J}^{(M)}(\omega)=\left(\frac{\mu}{\hbar}\right)^2 g_M^2(\omega)$, with $g_M$ defined in Eq. \eqref{eq:gM}, can be obtained from the power absorbed within the volume of the sphere when it is excited by a nearby dipole and derived in \cite{kim_classical_1988,francs_fluorescence_2008}:
\begin{widetext}
\begin{align}
\frac{\mathcal{J}^{(M)}_\parallel \left( \omega, d \right)}{\mathcal{J}_0} &=
-\frac{3}{4}  \left( \frac{k_\omega}{k_r} \right)^3 
 \sum_{n=1}^{\infty}\left(2n+1\right) \times  \left[\left|\frac{\zeta_n'(k_\omega d)}{k_\omega d}\right|^2\left(\operatorname{Re}\left(B_n\right)+\left|B_n\right|^2\right)+\left|h_n^{(1)}(k_\omega d)\right|^2\left(\operatorname{Re}\left(A_n\right)+\left|A_n\right|^2\right)\right], \\
\frac{\mathcal{J}^{(M)}_\perp \left( \omega, d \right)}{\mathcal{J}_0} &= -\frac{3}{2} \left( \frac{k_\omega}{k_r} \right)^3 \sum_{n=1}^{\infty} n(n+1)(2 n+1)\left|\frac{h_n^{(1)}(k_\omega d)}{k_\omega d}\right|^2\left(\operatorname{Re}\left(B_n\right)+\left|B_n\right|^2\right).
\end{align}
\end{widetext}

The quantity $\Gamma(\omega;\rb_a)$, defined in \eqref{eq:gequiv1}, can be derived by evaluating the imaginary part of the dyadic Green function, which has been analytically evaluated in \cite{kim_classical_1988}:

\begin{multline}
\frac{\Gamma_\parallel(\omega, d) }{2\pi \mathcal{J}_0} =  
 \left( \frac{k_\omega}{k_r} \right)^3+ \frac{3}{4} \left( \frac{k_\omega}{k_r} \right)^3 \times     \\  \Re \sum_{n=1}^{\infty}\left(2n+1\right)   \left[B_n\left(\frac{\zeta_n^{\prime}(k_\omega d)}{k_\omega d}\right)^2+A_n\left(h_n^{(1)}(k_\omega d)\right)^2\right],
\end{multline}

\begin{multline}
\frac{\Gamma_\perp(\omega, d) }{2\pi \mathcal{J}_0} =  
 \left( \frac{k_\omega}{k_r} \right)^3  \times  \\
\left\{1+\frac{3}{2} \operatorname{Re} \sum_{n=1}^{\infty} n(n+1)(2 n+1) B_n\left[\frac{h_n^{(1)}(k_\omega d)}{k_\omega d}\right]^2\right\}.
\end{multline}

\section{Resonance frequencies and fractional bandwidths of a small sphere}
\label{sec:Resonances}

In this appendix, we present closed-form expressions for the resonance frequencies of a dielectric sphere, considering both metal (Drude) and dielectric (Debye) media in the limit of small particle size.

In the limit $  k_p a  \rightarrow 0$, the resonance frequencies of a Drude sphere with small material losses are approximated by \cite{forestiere_resonance_2020}:

\begin{equation}
     \frac{\omega_n}{\omega_p} \approx   \sqrt{\frac{n}{2n+1}} \left[1 - \frac{(n+1)}{(3+2 n)(4 n^2-1)}  (k_p a)^2 \right], 
     \label{eq:PlasmonicRes}
\end{equation}
where $n$ denotes the multipolar order: $n=1$ (dipole), $n=2$ (quadrupole), $n=3$ (octupole), and so on.

The “fractional” bandwidth is the
bandwidth of a resonance divided by its center frequency. A common definition is the 3 dB
fractional bandwidth, which refers to the frequency range
within which the scattered power is within 3 dB of its maximum value, meaning the range of frequencies over which
the signal power is at least half of its peak value. The approximate 3 dB fractional bandwidth of a plasmonic mode is \cite{forestiere_resonance_2020}:
\begin{equation}
       \texttt{FWB}_n = \frac{\nu}{\omega_n} +  \frac{ \left( n + 1 \right) \left( 2 n  + 1 \right)}{ n  \left[ \left( 2n + 1 \right)!! \right]^2}  \left( \frac{a \omega_n }{c } \right)^{2n+1}.
    \label{eq:Qlsphere}
\end{equation}
which is the sum of the contribution from the material losses (first term) and radiation losses (second term).

In the limit $ k_c a  \rightarrow 0$, the resonance frequencies for a Debye sphere with negligible material losses are provided by the following expressions for magnetic type (H-type) and electric type (E-type) modes, respectively \cite{forestiere_resonance_2020}:

\begin{subequations}
\begin{align}
    \label{eq:DielResTE}
    \frac{\omega_{n,\ell}^{(H)}}{\omega_c} &=  z_{n-1,\ell}\left[ 1 + \frac{1}{2} \frac{2n+1}{2n-1} \, {(k_c  a)^2} \right], \\
    \label{eq:DielResTM}
    \frac{\omega_{n,\ell}^{(E)}}{\omega_c} &=  z_{n,\ell}\left[ 1 + \frac{1}{2} \frac{n+2}{n} \, {(k_c a)^2} \right], 
\end{align} 
\end{subequations}
where $z_{n,\ell}$ denotes the $\ell$-th zero of the spherical Bessel function $j_n$, $n$ represents the multipolar order of the resonance, and the index \(\ell\) indicates the number of maxima along the radial direction of the magnitude of the mode spatial distribution within the sphere. The \(H\)-type modes are characterized by a zero radial electric field and include magnetic dipole (\(n=1 \)), quadrupole (\(n=2 \)), and octupole (\(n=3 \)) modes of order $\ell=1,2,\ldots$.   The \(E\)-type modes are distinguished by a spatial distribution of the mode with vanishing radial magnetic field.  Examples include electric toroidal dipole (\(n=1\)), toroidal quadrupole (\(n=2\)) and toroidal octupole (\(n=3\)) modes of different order $\ell=1,2,\ldots$. 
The approximate 3 dB fractional bandwidths of the dielectric modes are:
\begin{subequations}
    \begin{align}
    \texttt{FBW}_{n,\ell}^{(H)} &\approx \tau \omega_{n,\ell}^{(H)} +  \frac{2 }{ \left[   z_{n-1,\ell}  \left( 2n-1\right)!! \right]^2}   \left( \frac{a \omega_{n,\ell}^{(H)} }{c} \right)^{2n+1}.
    \label{eq:QmqsTE} \\
    \texttt{FBW}_{n,\ell}^{(E)}  &\approx  \tau \omega_{n,\ell}^{(E)} + \frac{2}{  \left[  n \, z_{n,\ell} \left( 2n-1\right)!! \right]^2}   \left( \frac{a \omega_{n,\ell}^{(E)} }{c} \right)^{2n+3}.
    \label{eq:QmqsTM}
    \end{align}
\end{subequations}
which again consist of contributions from material losses (first term) and radiation losses (second term).
Further details regarding these resonances in dielectric spheres can be found in Refs.  \cite{forestiere_resonance_2020,forestiere_first-principles_2024}.

\section{Damping rate in the Born-Markov approximation }
\label{sec:QME}

The one-sided Fourier transform of the correlation function $\Sigma(\omega)=\int_0^\infty dt C(t) e^{i\omega t}$, i.e., the so-called spectral function, provides the fundamental parameters of the Lindblad master equation for the quantum emitter (e.g. \cite{h_p_breuer_and_f_petruccione_theory_2002}). The spectral function is expressed as $\Sigma(\omega)=(\hbar/\mu)^2[\frac{1}{2}\gamma (\omega) + i S(\omega)]$ where $\gamma(\omega)$ is the damping rate and $S(\omega)$ is the Lamb shift of the quantum emitter due to the interaction with the electromagnetic environment, in the Lindbladian approximation. Applying the relation $\int_0^\infty dt e^{i\omega t}=\pi \delta(\omega)+i\mathcal{P}\left(\frac{1}{\omega}\right)$ (e.g., \cite{w_heitler_quantum_1984})  to \eqref{eq:corr2}
we obtain for the damping rate
\begin{equation}
\gamma(\omega)=2\pi[\mathcal{J}^{(M)}_{td}(\omega)+\mathcal{J}_{td}^{(S)}(\omega)],
\end{equation}
and for the Lamb shift
\begin{equation}
    S(\omega)=\mathcal{P}\int_{-\infty}^{+\infty} d\omega' \frac{[\mathcal{J}^{(M)}_{td}(\omega')+\mathcal{J}_{td}^{(S)}(\omega')]}{\omega-\omega'},
\end{equation}
where $\mathcal{P}$ indicates the Cauchy principal value. The damping rate due to the interaction of the quantum emitter with the overall electromagnetic environment is the sum of the damping rate $\gamma_M(\omega)=2\pi\mathcal{J}^{(M)}_{td}(\omega)$ introduced by the medium-assisted bath and the damping rate $\gamma_S(\omega)=2\pi\mathcal{J}^{(S)}_{td}(\omega)$ introduced by the scattering-assisted bath. Analogously, the Lamb shift is the sum of the Lamb shifts resulting from each individual bath. The damping rate and the Lamb-shift account for both the medium-assisted and the scattering-assisted baths as a single one~\cite{potts2024qthermo}, so $\gamma(\omega)=2\pi\mathcal{J}_{eff}(\omega)$ and $S(\omega)=\mathcal{P}\int_{-\infty}^{+\infty} d\omega' \frac{\mathcal{J}_{eff}(\omega')}{\omega-\omega'}$. Note also that this holds true irrespective of the definition of the effective spectral density.

\section{Details on the MPS approach}
\label{sec:DetMPS}
The numerical approach employed to simulate the nonequilibrium dynamics of the composite system makes use of standard matrix product states representation \cite{schollwock_density-matrix_2011} of the composite ket state $\ket{\psi(t)}$. In fact, the simulation is performed by means of a single quantum emitter degree of freedom and a system of $N_{max}=400$ discrete bosonic oscillators, arranged in consecutive pairs of negative and positive frequencies. Each bosonic degree of freedom is allowed to have a maximum local dimension of the Hilbert space $N_{\text{ph}}=2$. To simulate dynamics, for any given time interval $[t,t+\mathrm{d}t]$ we employed a standard representation of the unitary operator that uses matrix product operators \cite{Zaletel2015}. The latter has proven suitable for studying long-range Hamiltonian operators and is implemented in the ITensor library \cite{itensor22}. To achieve our results, we set the maximum bond dimension of the simulation $D=200$. We also adopted a truncation error $\delta=10^{-12}$ and a $\mathrm{d}t=5\cdot 10^{-4} \omega^{-1}_\text{p}$. These choices allow for a good convergence of numerical results for the range of coupling strengths investigated and the frequency cutoff set in the main text.

\end{document}